\documentclass[aps,prb,amsmath,amssymb,10pt,superscriptaddress,twocolumn,color,epsfig,graphicx,bm]{revtex4-1}

\usepackage{color}
\usepackage{graphicx}
\usepackage{graphics}
\usepackage{epsfig}
\usepackage{amsmath}
\usepackage{array}
\usepackage{amssymb}
\usepackage{graphicx}      
\usepackage{dcolumn}       
\usepackage{bm}            
\usepackage{array}
\usepackage{booktabs}
\usepackage{ulem}

\usepackage[linktocpage=true,
  colorlinks=true, 
  pdfborder={0 0 0},
  linkcolor=blue,
  citecolor=red,
  filecolor=yellow,
  urlcolor=blue,
  bookmarks,
  pdfauthor={},
]{hyperref}

\newcommand{\tokyo}{Department of Applied Physics, The University of Tokyo, Tokyo 113-8656, Japan}

\begin{document}

\title{
Structural stability and energy levels of carbon-related defects in amorphous SiO$_2$ and its interface with SiC
}

\author{Yu-ichiro Matsushita}   
\affiliation{Laboratory for Materials and Structures, Institute of Innovative Research, Tokyo Institute of Technology, Yokohama 226-8503, Japan}
\affiliation{\tokyo}
\author{Atsushi Oshiyama}         
\affiliation{\tokyo}
\affiliation{Institute of Materials and Systems for Sustainability, Nagoya University, Nagoya 464-8601, Japan}

\date{\today}

\begin{abstract}
We report the density-functional calculations that systematically clarify the stable forms of carbon-related defects and their energy levels in amorphous SiO$_2$ using the melt-quench technique in molecular dynamics. Considering the position dependence of the O chemical potential near and far from the SiC/SiO$_2$ interface, we determine the most abundant forms of carbon-related defects: Far from the interface, the CO$_2$ or CO in the internal space in SiO$_2$ is abundant and they are electronically inactive; near the interface, the carbon clustering is likely and a particular mono-carbon defect and a di-carbon defect induce energy levels near the SiC conduction-band bottom, thus being candidates for the carrier traps. 
\end{abstract}

\pacs{~}
\maketitle

\section{Introduction}
Power electronics is one of fundamental technologies to realize our sustainable society. In this context, many researchers have paid attention to wide-gap semiconductors such as silicon carbide (SiC). The breakdown electric field of SiC is 10 times larger and its thermal conductance is 4 times larger than those of Si\cite{Kimoto,Baliga}. Furthermore, as one of the good advantages, amorphous SiO$_2$ is easily formed on the SiC surface upon its thermal oxidation and utilized as an insulator in metal oxide semiconductor (MOS) devices. However, even in the commercially available SiC-MOS devices, high density of interface levels (Dit) \cite{Yoshioka} and/or near interface traps (NIT) at the SiC/SiO$_2$ structure is a serious problem\cite{Tilak,Chow,Noborio}. 

To solve it, a lot of theoretical and experimental efforts have been done: Reported candidates for the Dit or NIT are carbon clusters\cite{Bassler, Gali1,Gali2,Gali3,Pantelides,Shiraishi1,Shiraishi2,Gavrikov,Pantelides1,Pantelides2,Pasquarello1,Pasquarello2}, oxygen interstitials\cite{Ono1,Ono2}, stacking faults\cite{Matsushita_Dit}, silicon interstitials in SiO$_2$\cite{Hijikata,Hijikata1}, SiC distortion induced by SiO$_2$\cite{Hirai,Shiraishi2}, and intrinsic defects of SiO$_2$\cite{Afanasev}.

In thermal oxidation of SiC, oxygen molecules or atoms diffuse in the SiO$_2$ film and react with Si atoms near the SiC/SiO$_2$ interface to form the SiO$_2$ bond network. During this process, C atoms near the SiC/SiO$_2$ interface are thought to be ejected as CO or CO$_2$ molecules and diffuse out eventually. However, substantial portion of the C atoms is expected to remain near the interface or in the SiO$_2$ films, presumably acting as Dit or NIT: In fact, Afanas'ev {\it et al.} discussed the existence of carbon clusters near the interface based on internal photoemission (IPE) spectroscopy\cite{Afanasev,Bassler}; some other groups detected the carbon-related defects (C defects) near the interface using secondary-ion-mass spectrometry (SIMS) and discussed the relation between the C defects and electron mobility\cite{Kobayashi,Kobayashi2}; from the theoretical side, density-functional calculations were done for SiC/SiO$_2$ interfaces and carbon-related defects (C-defects) are discussed \cite{Gali1,Gali2,Gali3,Pasquarello1,Pasquarello2}.


However, microscopic identification of C-defects in SiO$_2$ film has not been achieved yet. Works in the past have focused only on limited selections of C defects which are conjectured from empirical knowledge or educated guess, and then examined their properties. Systematic examination of the stability and the electronic structures of C defects is lacking. In this paper, we first prepare dozens of amorphous SiO$_2$ samples which contain carbon atoms by {\it ab-initio} molecular-dynamics simulations, then examine stability of various C defects based on the calculated formation energies, and clarify the energy levels induced by each C defect. We find that energetics of C defects is sensitive to chemical potentials of participating elements and thus the abundant forms of C defects depend on the position from the SiC/SiO$_2$ interface. We thus identify plausible C defects which are responsible for Dit and NIT.


\section{Calculations}
To explore abundant forms of C defects in amorphous SiO$_2$, we have performed melt-quench simulations based on Car-Parrinello Molecular Dynamics (CPMD) \cite{CPMD} in which SiO$_2$ samples containing C atoms are heated to liquids and then are quenched to local minimum structures. The obtained 45 samples are representatives of the C defects in amorphous SiO$_2$. In the actual simulations, we use our real-space scheme \cite{RSCPMD} based on the density functional theory (DFT) \cite{HK}, being implemented as RSDFT code \cite{RSCPMD1,RSCPMD2}. The grid spacing in RSDFT scheme is taken to be 0.19 \AA\ corresponding to the 840 eV cutoff energy in the plane-wave basis set. All simulations are done within the Perdew-Burke-Ernzerhof (PBE) functional\cite{PBE} for the exchange-correlation energy. A supercell containing 26 SiO$_2$ units and one or two C atoms is used and Brillouin zone (BZ) integration is performed with the $\Gamma$-point sampling. The ionic temperature is controlled by Nos\'e-Hoover thermostat \cite{Nose} in the constant NVT MD simulations. Samples are heated from 1000 K to 4500 K with the rate of 100 K/ps to be melt. After heating, we have quenched the samples to 1600 K with cooling rate of 10 K/ps. By this procedure, abundant forms of C defects are obtained without any empirical knowledge.

The formation energies of the C defects are calculated by embedding the obtained structures in the $\alpha$-quartz supercell and then optimizing the geometries. This is to eliminate the energy cost for the surrounding amorphous structure which depends on each sample, and to obtain the proper formation energy of each C defect on equal footing. The electron energy levels induced by a C defect are defined as the Fermi level positions in the gap at which the formation energies of the different charge states become equal (charge transition level \cite{Pasquarello2,car,Kumagai}). 
The actual computations in this part have been performed by using the Vienna $ab$ $initio$ simulation package (vasp)\cite{vasp}. We have used the cutoff energy of 400 eV in the plane-wave-basis set. The projector augmented wave (PAW) pseudopotentials \cite{PAW} with PBE functional have been used. The structural optimization has been done with a tolerance of the remaining forces less than $10^{-1}$ eV\AA$^{-1}$. When we calculate electron energy levels, we have used a hybrid functional of Hyde-Scuseria-Ernzerhof (HSE06)\cite{Hyde, Kresse, matsushita} in which Hartree-Fock mixing ratio of 0.5 and the screening parameter $\omega$ of 0.2 (au) are used to reproduce the band gap of SiO$_2$.

\section{Results and discussion}
\subsection{Mono carbon defects}

\begin{figure}
\includegraphics[width=1.0\linewidth]{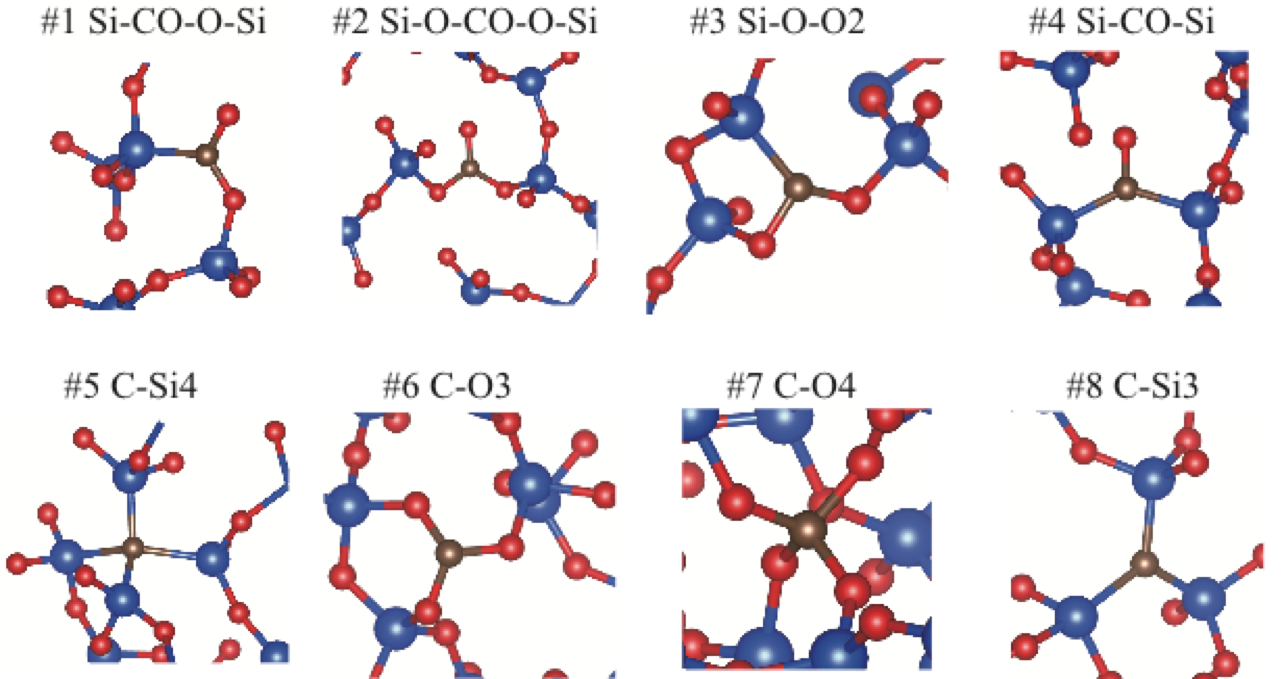}
\caption{(Color online) Atomic structures of mono-carbon defects obtained by the melt-quench method. Red, blue, and brown balls depict the Si, O, and C atoms.}
\label{Fig1}
\end{figure}

We start with mono-carbon defects. We prepare 25 samples which contain a single C atom and the different number of O atoms along with 26 SiO$_2$ units in a simulation cell, and performed the melt-quench simulations. The structure most frequently observed in the simulation is a CO molecule floating in a cavity of SiO$_2$ without making any chemical bonds with the SiO$_2$ network. We have then found 8 different structures of the C defects bonding with the SiO$_2$ network, as shown in Fig.~\ref{Fig1}: The C atom takes three (defects labeled as \#1, \#2,\#3, \#4, \#6 and \#8) or four (defects labeled as \#5 and \#7) coordination number; a dangling O appears in the defects \#1, \#2 and \#4.



\begin{figure}
\includegraphics[width=0.9\linewidth]{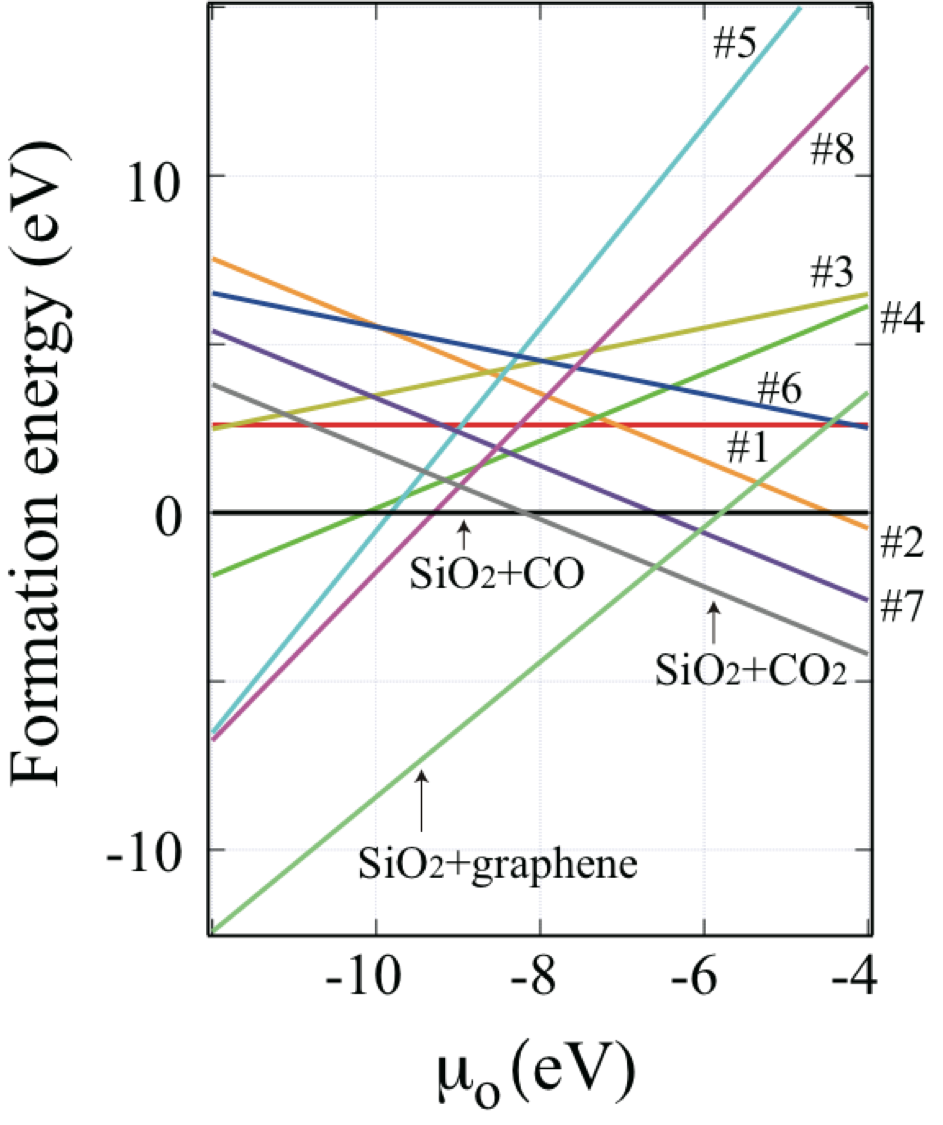} 
\caption{(Color online) Formation energy of each mono-carbon defect shown in Fig.~\ref{Fig1} as a function of oxygen chemical potential $\mu_{\rm O}$. The formation energies of CO, CO$_2$ and graphene in SiO$_2$ are also shown.}
\label{Fig2}
\end{figure}

\begin{figure*}
\includegraphics[width=0.8\linewidth]{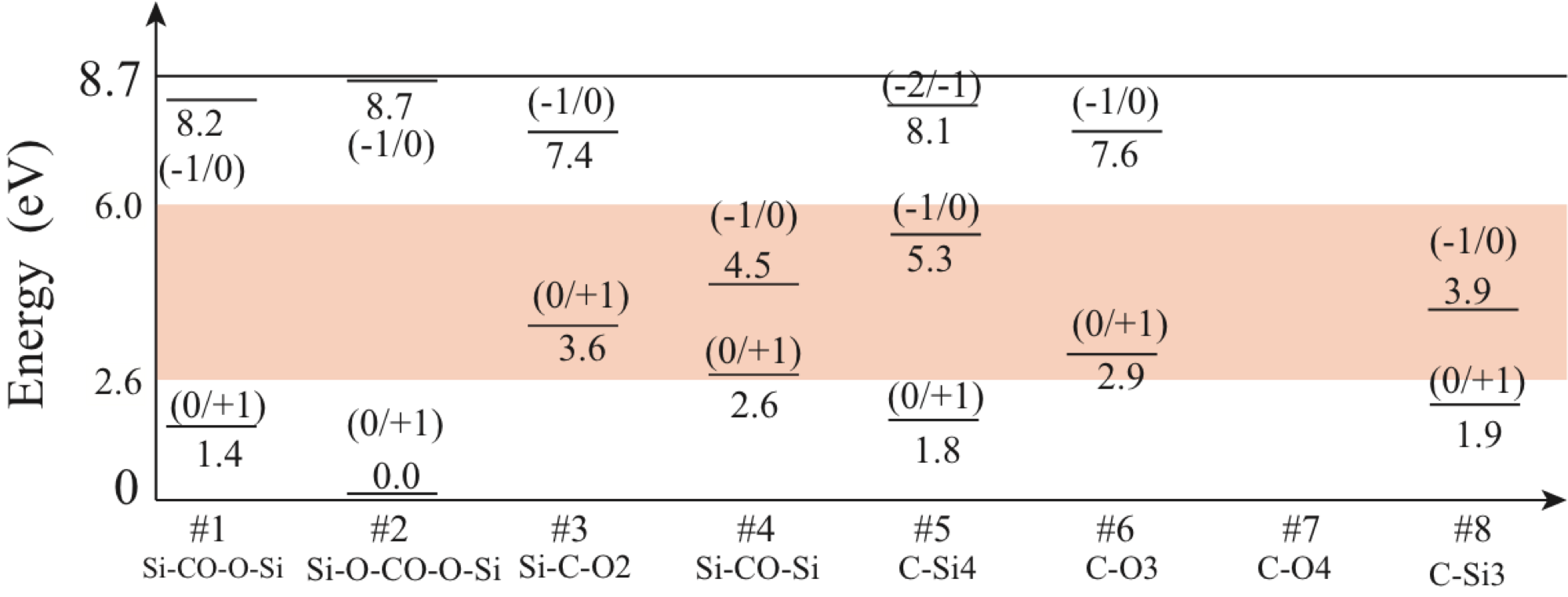}
\caption{(Color online) Electron energy levels (charge transition levels) of each mono-carbon defect shown in Fig.~\ref{Fig1} obtained by the HSE functional, in which the calculated SiO$_2$ gap is 8.7 eV. Orange area represents the SiC band-gap region. The band offset refers to the experimental value of Ref.~\onlinecite{offset}.}
\label{Fig3}
\end{figure*}

The formation energy $F$ of the C-defect is defined as 
\[
F = E_d - E_0 - N_{\rm Si}\mu_{\rm Si} - N_{\rm C}\mu_{\rm C}-N_{\rm O}\mu_{\rm O} , 
\]
where $E_d$ ($E_0$) is the total energy of the simulation cell with (without) the C-defect, and $N_{\rm A}$ and $\mu_{\rm A}$ are the number of atoms in the simulation cell and the chemical potential of the element A. Considering the relations, $\mu_{\rm Si} + 2 \mu_{\rm O} = E_{\rm SiO_2}$ and $\mu_{\rm C} + \mu_{\rm O} = E_{\rm CO}$ with $E_{\rm SiO_2}$ and $E_{\rm CO}$ being the total energies of SiO$_2$ and a  CO molecule, respectively, the formation energy is expressed as a function of oxygen atom, e.g, $\mu_{\rm O}$. The formation energy thus obtained is shown in Fig.~\ref{Fig2}. The range of $\mu_{\rm O}$ in Fig.~\ref{Fig2} includes the value for SiO$_2$ ($-$9 eV corresponding to the O-poor condition) and the value for an O$_2$ molecule ($-$4 eV: the O-rich condition). As a reference, we have also calculated the total energy of SiO$_2$ with a graphene sheet, modeling a situation where C atoms cluster in some forms. 


In the O-rich region ( $-$ 6 eV $\leq \mu_{\rm O} \leq$ $-$ 4 eV), we have found that the formation energy of the CO$_2$ molecule is the lowest and the C defect \#7, where 4 O atoms surround the C atom, is the 2'nd lowest. The C defect \#2 and the CO molecule follow. In contrast, in the O-poor region graphene in SiO$_2$ is the lowest, being indicative that C atoms tend to aggregate in the O-poor region. The C defects \#8 and \#5, where the C atom is surrounded by Si atoms alone, are relatively low in the formation energy. The C defect \#4, in which the C atom is bonded with two Si and one O atoms, follows them. These defects thus are likely to exist under imperfect oxidation of SiC. In the realistic situation during the oxidation of SiC, $\mu_{\rm O}$ is likely to have a position dependence along the direction perpendicular to the interface: Near the SiC/SiO$_2$ interface $\mu_{\rm O}$ should be lower (O poor) compared with that far from the interface. Our results thus indicate that carbon clusters are easily formed (the defects \#8, \# 5, and \#4 may also be possible) near the interface. On the other hand, far from the interface, CO$_2$ molecules should be majority with the small amount of the defect \#7 and \#2.


Figure \ref{Fig3} shows calculated energy level $(q/q^{\prime})$ at which the formation energies of the charge states $q$ and $q^{\prime}$ are equal. It is of note that neither CO$_2$ nor CO induces the energy level in the gap region of SiC. From the energetics viewpoint, the C defect \#7 (O-rich) or \#5 and \#8 (O poor) are expected to be abundant, and \#2 (O-rich) or \#4 (O-poor) follows.
We have found that the defects \#4, \#5, and \#8 induce energy levels in the energy gap of SiC: The defect \#5 induces an energy level $(-1/0)$ near the conduction band bottom whereas the defect \#4 and \#8 do a deep level $(-1/0)$. It is noteworthy that the defect \#7 and \#2 which are expected to exist in the O-rich region induces no levels in the gap. The defects \#4, \#5, and \#8 which are expected to exist in the O-poor region are important near the SiO$_2$/SiC interface. In particular, the defect \#5 which induces a level near the conduction band is a strong candidate for the carrier trap. 

\subsection{Intrinsic defects in SiO$_2$}

\begin{figure}
\includegraphics[width=0.7\linewidth]{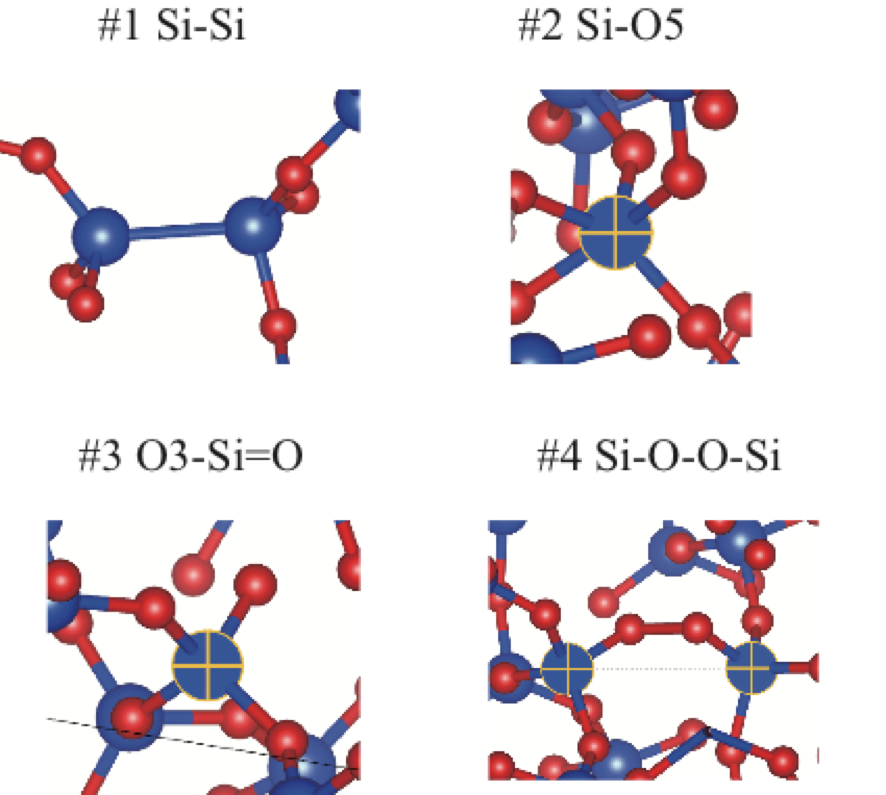}
\caption{(Color online) Atomic structures of intrinsic defects in SiO$_2$. Red and blue balls depict the Si and O atoms. In the \#2 defect, the marked Si atom is 5-fold coordinated. In the \#3 defect, the marked Si atom has a dangling O atom.}
\label{Fig4}
\end{figure}
\begin{figure}
\includegraphics[width=0.8\linewidth]{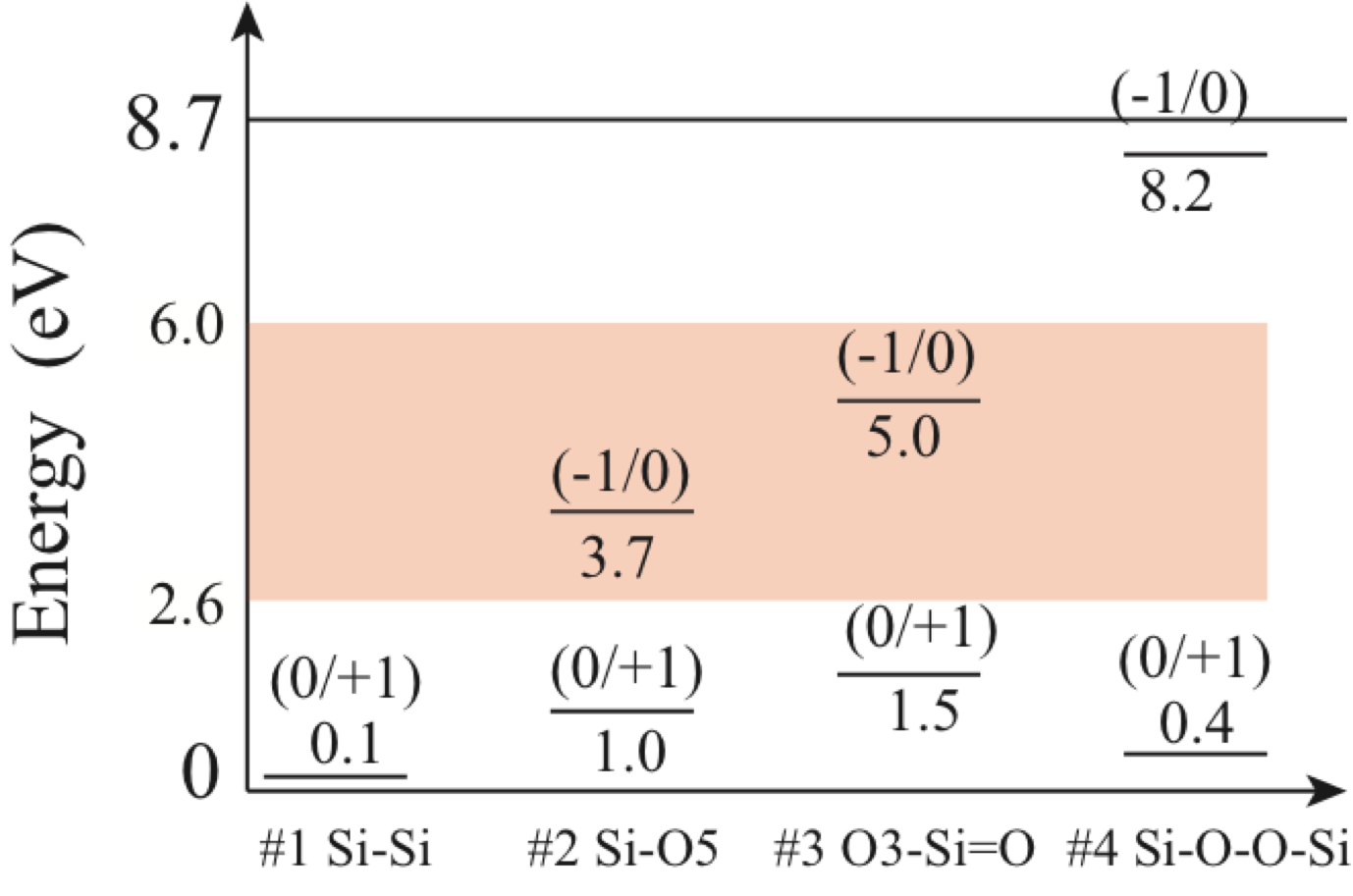}
\caption{(Color online) Electron energy levels (charge transition levels) of each intrinsic defect in SiO$_2$ shown in Fig.~\ref{Fig4} obtained by the HSE functional. Orange area represents the SiC band-gap region. The band offset refers to the experimental value of  Ref.~\onlinecite{offset}.}
\label{Fig5}
\end{figure}
We also discuss the intrinsic defects of SiO$_2$. In the above samples generated by the melt-quench method, we have observed 4 intrinsic defects in amorphous SiO$_2$ as shown in Fig.~\ref{Fig4}: Oxygen vacancy (labeled as \#1), 5 fold-coordinated Si atom (labeled as \#2), dangling Si=O bond (labeled as \#3), and peroxide (labeled as \#4). We have calculated the electronic energy levels of them in Fig.~\ref{Fig5}. All the defect structures cause several energy levels in the SiO$_2$ energy gap. Especially, \#2 and \#3 defects induce deep levels in the gap of SiC. 


\begin{figure}
\includegraphics[width=0.9\linewidth]{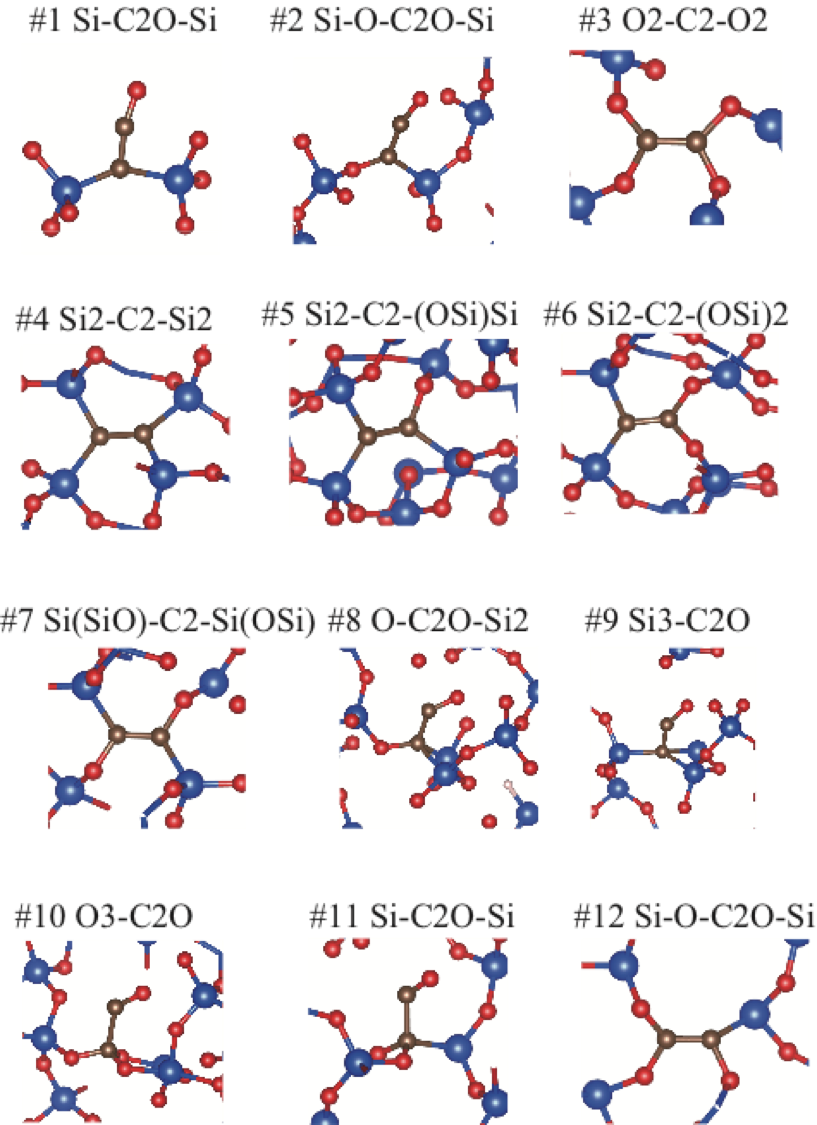}
\caption{(Color online) Atomic structures of di-carbon defects obtained by the melt-quench method. Red, blue, and brown balls depict the Si, O, and C atoms, respectively.}
\label{Fig6}
\end{figure}

\subsection{Di carbon defects}
We have then explored di-carbon defects by the melt-quench technique described above. We have prepared 20 amorphous SiO$_2$ samples consisting of 24 SiO$_2$ units with 2 carbon atoms. We have found that 2 carbon atoms are mostly relaxed to the mono-carbon structures that we have found above. In order to explore di-carbon structures, we use the blue moon method\cite{BM} in which we impose a constraint that the distance between 2 carbon atoms is kept to be 1.3 {\AA}. Consequently, we have found 12 distinct di-carbon-related defects in total, as shown in Fig.~\ref{Fig6}. 


The defect structures of the C atoms are categorized into 2 groups: line shape and ethylene-like shape: \#1, \#2, \#8, \#9, \#10, and \#11 are of the line shape structure, while \#3, \#4, \#5, \#6, \#7 and \#12 are of the ethylene-like structure. In the line-shape structures, one C atom is two-fold coordinated and the other C is three-fold (\#1, and \#2) or four-fold coordinated (\#8, \#9, \#10, and \#11). 


\begin{figure}
\includegraphics[width=0.9\linewidth]{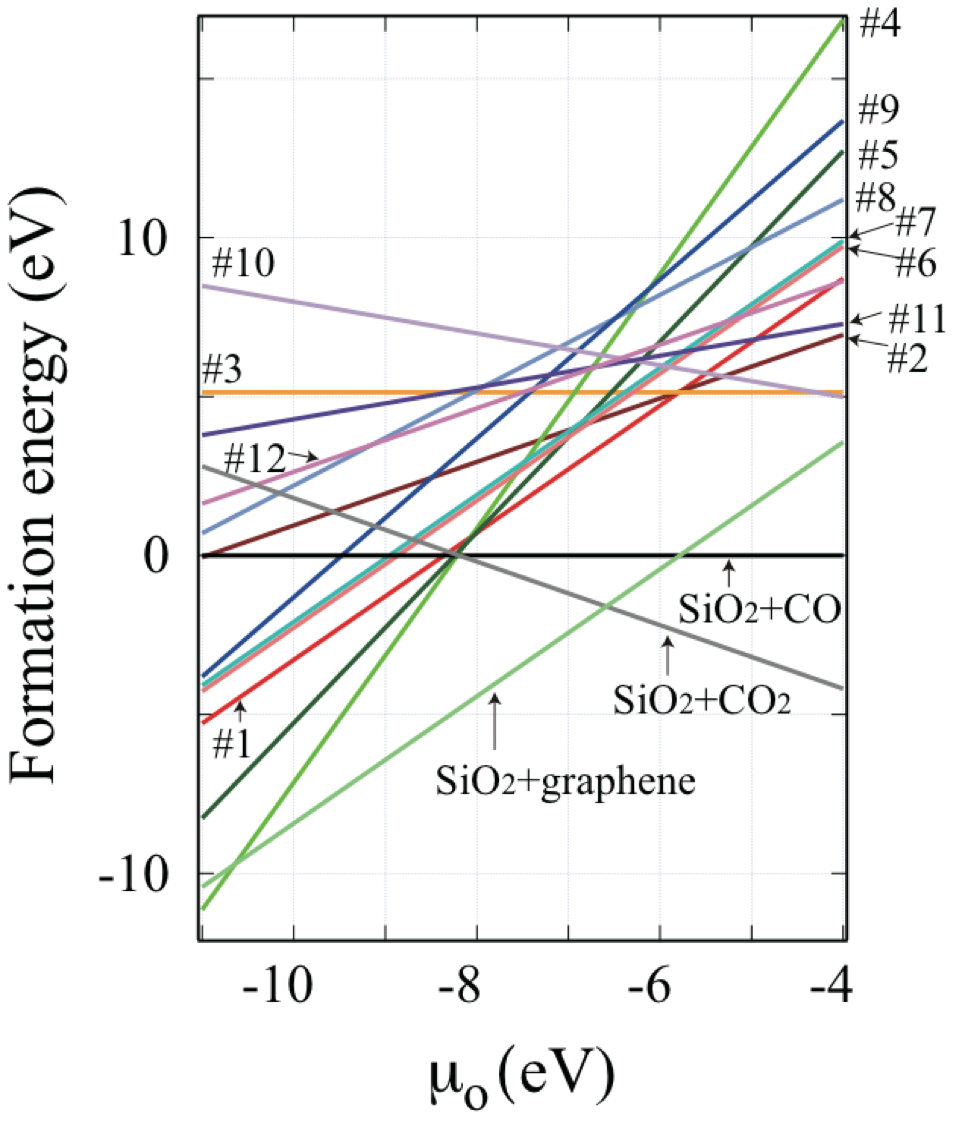}
\caption{(Color online) 
Formation energy of each di-carbon defect in Fig.~\ref{Fig6} as a function of oxygen chemical potential $\mu_{\rm O}$. The formation energies of CO, CO$_2$ and graphene in SiO$_2$ are also shown.}
\label{Fig7}
\end{figure}

\begin{figure*}
\includegraphics[width=1.0\linewidth]{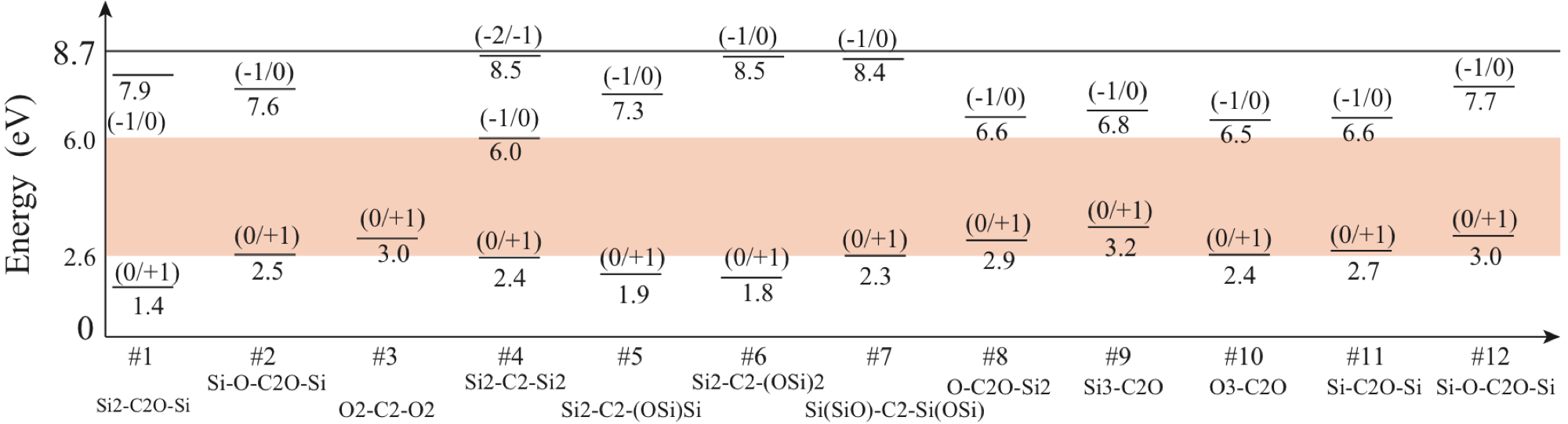}
\caption{(Color online) Electron energy levels (charge transition levels) of each di-carbon defect shown in Fig.~\ref{Fig6} obtained by the HSE functional. Orange area represents the SiC band-gap region. The band offset refers to the experimental value of Ref.~\onlinecite{offset}.}
\label{Fig8}
\end{figure*}

Figure \ref{Fig7} shows calculated formation energies of these di-carbon defects. In the O-rich region ranging from $\mu_{\rm O}$ = $-$7 to $-$4 eV, we have found that the CO$_2$ molecule is the lowest in the formation energy even when we extend our search to di-carbon defects. We have also found that the second lowest is the CO molecule and that the formation energies of the di-carbon defects in this O-rich region is higher by several eV or more (Fig.~\ref{Fig7}). This fact means that in the O-rich region the existence of the di-carbon defects is unlikely. Since oxygen is relatively rich in the SiO$_2$ region compared with the SiC/SiO$_2$ interface region, the CO$_2$ or the CO molecule in the cavity of SiO$_2$ is most abundant and the mono-carbon rather than the di-carbon defect exists as a minority. In contrast, in the O-poor region which may be the region near the SiC/SiO$_2$ interface, the generation of carbon clusters is the energetically favorable. However, interestingly the 2nd lowest structures are the di-carbon defects in the order of \#4, \#5, and \#1. In these di-carbon defects, C atoms are connected mainly with Si atoms. This feature is likely to appear under the imperfect oxidation and the present results clarify that this is energetically favorable. We have also performed ab-initio oxidation simulations considering SiC substrate explicitly and found the appearance of the di-carbon defects during the oxidation\cite{matsushita_arXiv}. The di-carbon formation is also reported in Refs.~\onlinecite{Shiraishi2, Akiyama}.


Figure \ref{Fig8} shows calculated energy level $(q/q^{\prime})$ induced by the di-carbon defects. As stated above, neither CO nor CO$_2$ in the cavity of SiO$_2$ induces the energy level in the gap region of SiC. From the energetics viewpoint, the defects \#4, \#5 and \#1 may be abundant in the O-poor region (near the interface) but the \#5 and \#1 does not induce the energy level in the gap. The defect \#4, however, induces the energy level just below the SiC conduction-band bottom, being a candidate for the carrier trap. 


\section{Conclusion}\label{conclusion}
On the basis of the density-functional theory, we have generated amorphous SiO$_2$ and extensively explored stable structures of carbon-related defects using the melt-quench scheme, and then clarified the induced electron energy levels (charge transition levels) near the energy gap of SiC. The stability of the defects have been clarified in terms of the oxygen chemical potential which corresponds to the oxygen abundance near and far from the SiO$_2$/SiC interface.

In the region with the O-poor condition, i.e., near the SiO$_2$/SiC interface where the oxidation may be incomplete, it is found that carbon clustering is energetically favorable. In particular, a di-carbon defect in the form of Si$_2$-C$_2$-Si$_2$ (the defect \#4 in Fig.~\ref{Fig6}) is the most stable carbon-related defect and induces an energy level just below the conduction-band bottom of SiC, being a strong candidate for the electron trap. We have also found that a mono-carbon defect in the form of C-Si$_4$ (defect \#5 in Fig.~\ref{Fig1}) is energetically favorable and induces a trap level near the conduction-band bottom. 

In the region with the O-rich condition, i.e., far from the SiO$_2$/SiC interface where the oxidation has been done enough, we have generally found that the most stable form of carbon is CO or CO$_2$ in the cavity of SiO$_2$. We have found that these forms induce no electronic levels in the gap region of SiC. We have also found two mono-carbon defects with relatively low formation energies. They have the shapes of the Si-O-CO-O-Si (the defect \#2 in Fig.~\ref{Fig2}) or the C-O4 (the defect \#7). We have found that these defects are electronically inactive.

\

\begin{acknowledgments}
Computations has been done with the resources of the Oak-forest PACKS provided by Multidisciplinary Cooperative Research Program in Center for Computational Sciences, University of Tsukuba and the K computer provided by the RIKEN Advanced Institute for Computational Science through the HPCI System Research project (Project ID: hp180227). We acknowledge the support from JSPS Grant-in-Aid for Scientific Research(A) (Grant Nos. 18H03770 and 18H03873).
\end{acknowledgments}

\bibliographystyle{apsrev4-1}
\bibliography{RP180358}

\begin{thebibliography}{45}%
\makeatletter
\providecommand \@ifxundefined [1]{%
 \@ifx{#1\undefined}
}%
\providecommand \@ifnum [1]{%
 \ifnum #1\expandafter \@firstoftwo
 \else \expandafter \@secondoftwo
 \fi
}%
\providecommand \@ifx [1]{%
 \ifx #1\expandafter \@firstoftwo
 \else \expandafter \@secondoftwo
 \fi
}%
\providecommand \natexlab [1]{#1}%
\providecommand \enquote  [1]{``#1''}%
\providecommand \bibnamefont  [1]{#1}%
\providecommand \bibfnamefont [1]{#1}%
\providecommand \citenamefont [1]{#1}%
\providecommand \href@noop [0]{\@secondoftwo}%
\providecommand \href [0]{\begingroup \@sanitize@url \@href}%
\providecommand \@href[1]{\@@startlink{#1}\@@href}%
\providecommand \@@href[1]{\endgroup#1\@@endlink}%
\providecommand \@sanitize@url [0]{\catcode `\\12\catcode `\$12\catcode
  `\&12\catcode `\#12\catcode `\^12\catcode `\_12\catcode `\%12\relax}%
\providecommand \@@startlink[1]{}%
\providecommand \@@endlink[0]{}%
\providecommand \url  [0]{\begingroup\@sanitize@url \@url }%
\providecommand \@url [1]{\endgroup\@href {#1}{\urlprefix }}%
\providecommand \urlprefix  [0]{URL }%
\providecommand \Eprint [0]{\href }%
\providecommand \doibase [0]{http://dx.doi.org/}%
\providecommand \selectlanguage [0]{\@gobble}%
\providecommand \bibinfo  [0]{\@secondoftwo}%
\providecommand \bibfield  [0]{\@secondoftwo}%
\providecommand \translation [1]{[#1]}%
\providecommand \BibitemOpen [0]{}%
\providecommand \bibitemStop [0]{}%
\providecommand \bibitemNoStop [0]{.\EOS\space}%
\providecommand \EOS [0]{\spacefactor3000\relax}%
\providecommand \BibitemShut  [1]{\csname bibitem#1\endcsname}%
\let\auto@bib@innerbib\@empty
\bibitem [{\citenamefont {Kimoto}\ and\ \citenamefont {Cooper}(2014)}]{Kimoto}%
  \BibitemOpen
  \bibfield  {author} {\bibinfo {author} {\bibfnamefont {T.}~\bibnamefont
  {Kimoto}}\ and\ \bibinfo {author} {\bibfnamefont {J.~A.}\ \bibnamefont
  {Cooper}},\ }\href@noop {} {\emph {\bibinfo {title} {Fundamentals of Silicon
  Carbide Technology}}}\ (\bibinfo {year} {2014})\BibitemShut {NoStop}%
\bibitem [{\citenamefont {Baliga}(1989)}]{Baliga}%
  \BibitemOpen
  \bibfield  {author} {\bibinfo {author} {\bibfnamefont {B.~J.}\ \bibnamefont
  {Baliga}},\ }\href@noop {} {\bibfield  {journal} {\bibinfo  {journal} {IEEE
  Electron Device Lett.}\ }\textbf {\bibinfo {volume} {10}},\ \bibinfo {pages}
  {455} (\bibinfo {year} {1989})}\BibitemShut {NoStop}%
\bibitem [{\citenamefont {Yoshioka}\ \emph {et~al.}(2015)\citenamefont
  {Yoshioka}, \citenamefont {Senzaki}, \citenamefont {Shimozato}, \citenamefont
  {Tanaka},\ and\ \citenamefont {Okumura}}]{Yoshioka}%
  \BibitemOpen
  \bibfield  {author} {\bibinfo {author} {\bibfnamefont {H.}~\bibnamefont
  {Yoshioka}}, \bibinfo {author} {\bibfnamefont {J.}~\bibnamefont {Senzaki}},
  \bibinfo {author} {\bibfnamefont {A.}~\bibnamefont {Shimozato}}, \bibinfo
  {author} {\bibfnamefont {Y.}~\bibnamefont {Tanaka}}, \ and\ \bibinfo {author}
  {\bibfnamefont {H.}~\bibnamefont {Okumura}},\ }\href@noop {} {\bibfield
  {journal} {\bibinfo  {journal} {AIP Advances}\ }\textbf {\bibinfo {volume}
  {5}},\ \bibinfo {pages} {017109} (\bibinfo {year} {2015})}\BibitemShut
  {NoStop}%
\bibitem [{Til()}]{Tilak}%
  \BibitemOpen
  \href@noop {} {}\bibinfo {note} {V. Tilak, Phys. Status Solidi A, {\bf 206},
  2391 (2009).}\BibitemShut {Stop}%
\bibitem [{Cho()}]{Chow}%
  \BibitemOpen
  \href@noop {} {}\bibinfo {note} {T. P. Chow, H. Naik, and Z. Li, Phys. Status
  Solidi A, {\bf 206}, 2478 (2009).}\BibitemShut {Stop}%
\bibitem [{Nob()}]{Noborio}%
  \BibitemOpen
  \href@noop {} {}\bibinfo {note} {M. Noborio, J. Suda, S. Beljakowa, M.
  Krieger, and T. Kimoto, Phys. Status Solidi A, {\bf 206}, 2374
  (2009).}\BibitemShut {Stop}%
\bibitem [{\citenamefont {Bassler}\ \emph {et~al.}(1997)\citenamefont
  {Bassler}, \citenamefont {Pensl},\ and\ \citenamefont {Afanas'ev}}]{Bassler}%
  \BibitemOpen
  \bibfield  {author} {\bibinfo {author} {\bibfnamefont {M.}~\bibnamefont
  {Bassler}}, \bibinfo {author} {\bibfnamefont {G.}~\bibnamefont {Pensl}}, \
  and\ \bibinfo {author} {\bibfnamefont {V.}~\bibnamefont {Afanas'ev}},\ }\href
  {\doibase https://doi.org/10.1016/S0925-9635(97)00074-5} {\bibfield
  {journal} {\bibinfo  {journal} {Diamond and Related Materials}\ }\textbf
  {\bibinfo {volume} {6}},\ \bibinfo {pages} {1472 } (\bibinfo {year}
  {1997})},\ \bibinfo {note} {proceeding of the 1st European Conference on
  Silicon Carbide and Related Materials (ECSCRM 1996)}\BibitemShut {NoStop}%
\bibitem [{\citenamefont {Knaup}\ \emph
  {et~al.}(2005{\natexlab{a}})\citenamefont {Knaup}, \citenamefont {De\'ak},
  \citenamefont {Frauenheim}, \citenamefont {Gali}, \citenamefont {Hajnal},\
  and\ \citenamefont {Choyke}}]{Gali1}%
  \BibitemOpen
  \bibfield  {author} {\bibinfo {author} {\bibfnamefont {J.~M.}\ \bibnamefont
  {Knaup}}, \bibinfo {author} {\bibfnamefont {P.}~\bibnamefont {De\'ak}},
  \bibinfo {author} {\bibfnamefont {T.}~\bibnamefont {Frauenheim}}, \bibinfo
  {author} {\bibfnamefont {A.}~\bibnamefont {Gali}}, \bibinfo {author}
  {\bibfnamefont {Z.}~\bibnamefont {Hajnal}}, \ and\ \bibinfo {author}
  {\bibfnamefont {W.~J.}\ \bibnamefont {Choyke}},\ }\href {\doibase
  10.1103/PhysRevB.71.235321} {\bibfield  {journal} {\bibinfo  {journal} {Phys.
  Rev. B}\ }\textbf {\bibinfo {volume} {71}},\ \bibinfo {pages} {235321}
  (\bibinfo {year} {2005}{\natexlab{a}})}\BibitemShut {NoStop}%
\bibitem [{\citenamefont {Knaup}\ \emph
  {et~al.}(2005{\natexlab{b}})\citenamefont {Knaup}, \citenamefont {De\'ak},
  \citenamefont {Frauenheim}, \citenamefont {Gali}, \citenamefont {Hajnal},\
  and\ \citenamefont {Choyke}}]{Gali2}%
  \BibitemOpen
  \bibfield  {author} {\bibinfo {author} {\bibfnamefont {J.~M.}\ \bibnamefont
  {Knaup}}, \bibinfo {author} {\bibfnamefont {P.}~\bibnamefont {De\'ak}},
  \bibinfo {author} {\bibfnamefont {T.}~\bibnamefont {Frauenheim}}, \bibinfo
  {author} {\bibfnamefont {A.}~\bibnamefont {Gali}}, \bibinfo {author}
  {\bibfnamefont {Z.}~\bibnamefont {Hajnal}}, \ and\ \bibinfo {author}
  {\bibfnamefont {W.~J.}\ \bibnamefont {Choyke}},\ }\href {\doibase
  10.1103/PhysRevB.72.115323} {\bibfield  {journal} {\bibinfo  {journal} {Phys.
  Rev. B}\ }\textbf {\bibinfo {volume} {72}},\ \bibinfo {pages} {115323}
  (\bibinfo {year} {2005}{\natexlab{b}})}\BibitemShut {NoStop}%
\bibitem [{\citenamefont {De\'ak}\ \emph {et~al.}(2007)\citenamefont {De\'ak},
  \citenamefont {Knaup}, \citenamefont {Hornos}, \citenamefont {Thill},
  \citenamefont {Gali},\ and\ \citenamefont {Frauenheim}}]{Gali3}%
  \BibitemOpen
  \bibfield  {author} {\bibinfo {author} {\bibfnamefont {P.}~\bibnamefont
  {De\'ak}}, \bibinfo {author} {\bibfnamefont {J.~M.}\ \bibnamefont {Knaup}},
  \bibinfo {author} {\bibfnamefont {T.}~\bibnamefont {Hornos}}, \bibinfo
  {author} {\bibfnamefont {C.}~\bibnamefont {Thill}}, \bibinfo {author}
  {\bibfnamefont {A.}~\bibnamefont {Gali}}, \ and\ \bibinfo {author}
  {\bibfnamefont {T.}~\bibnamefont {Frauenheim}},\ }\href
  {http://stacks.iop.org/0022-3727/40/i=20/a=S09} {\bibfield  {journal}
  {\bibinfo  {journal} {Journal of Physics D: Applied Physics}\ }\textbf
  {\bibinfo {volume} {40}},\ \bibinfo {pages} {6242} (\bibinfo {year}
  {2007})}\BibitemShut {NoStop}%
\bibitem [{Pan()}]{Pantelides}%
  \BibitemOpen
  \href@noop {} {}\bibinfo {note} {X. Shen and S. T. Pantelides, App. Phys.
  Lett., {\bf 98}, 053507 (2011).}\BibitemShut {Stop}%
\bibitem [{\citenamefont {Chokawa}\ \emph {et~al.}(2013)\citenamefont
  {Chokawa}, \citenamefont {Kato}, \citenamefont {Kamiya},\ and\ \citenamefont
  {Shiraishi}}]{Shiraishi1}%
  \BibitemOpen
  \bibfield  {author} {\bibinfo {author} {\bibfnamefont {K.}~\bibnamefont
  {Chokawa}}, \bibinfo {author} {\bibfnamefont {S.}~\bibnamefont {Kato}},
  \bibinfo {author} {\bibfnamefont {K.}~\bibnamefont {Kamiya}}, \ and\ \bibinfo
  {author} {\bibfnamefont {K.}~\bibnamefont {Shiraishi}},\ }\href@noop {}
  {\bibfield  {journal} {\bibinfo  {journal} {Materials Science Forum}\
  }\textbf {\bibinfo {volume} {740}},\ \bibinfo {pages} {469} (\bibinfo {year}
  {2013})}\BibitemShut {NoStop}%
\bibitem [{\citenamefont {Shiraishi}\ \emph {et~al.}(2014)\citenamefont
  {Shiraishi}, \citenamefont {Chokawa}, \citenamefont {Shirakawa},
  \citenamefont {Endo}, \citenamefont {Araidai}, \citenamefont {Kamiya},\ and\
  \citenamefont {Watanabe}}]{Shiraishi2}%
  \BibitemOpen
  \bibfield  {author} {\bibinfo {author} {\bibfnamefont {K.}~\bibnamefont
  {Shiraishi}}, \bibinfo {author} {\bibfnamefont {K.}~\bibnamefont {Chokawa}},
  \bibinfo {author} {\bibfnamefont {H.}~\bibnamefont {Shirakawa}}, \bibinfo
  {author} {\bibfnamefont {K.}~\bibnamefont {Endo}}, \bibinfo {author}
  {\bibfnamefont {M.}~\bibnamefont {Araidai}}, \bibinfo {author} {\bibfnamefont
  {K.}~\bibnamefont {Kamiya}}, \ and\ \bibinfo {author} {\bibfnamefont
  {H.}~\bibnamefont {Watanabe}},\ }\href@noop {} {\bibfield  {journal}
  {\bibinfo  {journal} {2014 IEEE International Electron Devices Meeting}\
  }\textbf {\bibinfo {volume} {14}},\ \bibinfo {pages} {538} (\bibinfo {year}
  {2014})}\BibitemShut {NoStop}%
\bibitem [{Gav()}]{Gavrikov}%
  \BibitemOpen
  \href@noop {} {}\bibinfo {note} {A. Gavrikov, A. Knizhnik, A. Safonov, A.
  Scherbinin, A. Bagaturyants, B. Potapkin, A. Chatterjee, and K. Matocha, J.
  Appl. Phys. {\bf 104}, 093508 (2008).}\BibitemShut {Stop}%
\bibitem [{\citenamefont {Wang}\ \emph {et~al.}(2007)\citenamefont {Wang},
  \citenamefont {Dhar}, \citenamefont {Wang}, \citenamefont {Ahyi},
  \citenamefont {Franceschetti}, \citenamefont {Williams}, \citenamefont
  {Feldman},\ and\ \citenamefont {Pantelides}}]{Pantelides1}%
  \BibitemOpen
  \bibfield  {author} {\bibinfo {author} {\bibfnamefont {S.}~\bibnamefont
  {Wang}}, \bibinfo {author} {\bibfnamefont {S.}~\bibnamefont {Dhar}}, \bibinfo
  {author} {\bibfnamefont {S.}~\bibnamefont {Wang}}, \bibinfo {author}
  {\bibfnamefont {A.~C.}\ \bibnamefont {Ahyi}}, \bibinfo {author}
  {\bibfnamefont {A.}~\bibnamefont {Franceschetti}}, \bibinfo {author}
  {\bibfnamefont {J.~R.}\ \bibnamefont {Williams}}, \bibinfo {author}
  {\bibfnamefont {L.~C.}\ \bibnamefont {Feldman}}, \ and\ \bibinfo {author}
  {\bibfnamefont {S.~T.}\ \bibnamefont {Pantelides}},\ }\href@noop {}
  {\bibfield  {journal} {\bibinfo  {journal} {Phys. Rev. Lett.}\ }\textbf
  {\bibinfo {volume} {98}},\ \bibinfo {pages} {026101} (\bibinfo {year}
  {2007})}\BibitemShut {NoStop}%
\bibitem [{\citenamefont {Wang}\ \emph {et~al.}(2001)\citenamefont {Wang},
  \citenamefont {DiVentra}, \citenamefont {Kim},\ and\ \citenamefont
  {Pantelides}}]{Pantelides2}%
  \BibitemOpen
  \bibfield  {author} {\bibinfo {author} {\bibfnamefont {S.}~\bibnamefont
  {Wang}}, \bibinfo {author} {\bibfnamefont {M.}~\bibnamefont {DiVentra}},
  \bibinfo {author} {\bibfnamefont {S.~G.}\ \bibnamefont {Kim}}, \ and\
  \bibinfo {author} {\bibfnamefont {S.~T.}\ \bibnamefont {Pantelides}},\
  }\href@noop {} {\bibfield  {journal} {\bibinfo  {journal} {Phys. Rev. Lett.}\
  }\textbf {\bibinfo {volume} {86}},\ \bibinfo {pages} {5946} (\bibinfo {year}
  {2001})}\BibitemShut {NoStop}%
\bibitem [{\citenamefont {Devynck}\ \emph
  {et~al.}(2011{\natexlab{a}})\citenamefont {Devynck}, \citenamefont
  {Alkauskas}, \citenamefont {Broqvist},\ and\ \citenamefont
  {Pasquarello}}]{Pasquarello1}%
  \BibitemOpen
  \bibfield  {author} {\bibinfo {author} {\bibfnamefont {F.}~\bibnamefont
  {Devynck}}, \bibinfo {author} {\bibfnamefont {A.}~\bibnamefont {Alkauskas}},
  \bibinfo {author} {\bibfnamefont {P.}~\bibnamefont {Broqvist}}, \ and\
  \bibinfo {author} {\bibfnamefont {A.}~\bibnamefont {Pasquarello}},\ }\href
  {\doibase 10.1103/PhysRevB.83.195319} {\bibfield  {journal} {\bibinfo
  {journal} {Phys. Rev. B}\ }\textbf {\bibinfo {volume} {83}},\ \bibinfo
  {pages} {195319} (\bibinfo {year} {2011}{\natexlab{a}})}\BibitemShut
  {NoStop}%
\bibitem [{\citenamefont {Devynck}\ \emph
  {et~al.}(2011{\natexlab{b}})\citenamefont {Devynck}, \citenamefont
  {Alkauskas}, \citenamefont {Broqvist},\ and\ \citenamefont
  {Pasquarello}}]{Pasquarello2}%
  \BibitemOpen
  \bibfield  {author} {\bibinfo {author} {\bibfnamefont {F.}~\bibnamefont
  {Devynck}}, \bibinfo {author} {\bibfnamefont {A.}~\bibnamefont {Alkauskas}},
  \bibinfo {author} {\bibfnamefont {P.}~\bibnamefont {Broqvist}}, \ and\
  \bibinfo {author} {\bibfnamefont {A.}~\bibnamefont {Pasquarello}},\ }\href
  {\doibase 10.1103/PhysRevB.84.235320} {\bibfield  {journal} {\bibinfo
  {journal} {Phys. Rev. B}\ }\textbf {\bibinfo {volume} {84}},\ \bibinfo
  {pages} {235320} (\bibinfo {year} {2011}{\natexlab{b}})}\BibitemShut
  {NoStop}%
\bibitem [{\citenamefont {Kirkham}\ and\ \citenamefont {Ono}(2016)}]{Ono1}%
  \BibitemOpen
  \bibfield  {author} {\bibinfo {author} {\bibfnamefont {C.~J.}\ \bibnamefont
  {Kirkham}}\ and\ \bibinfo {author} {\bibfnamefont {T.}~\bibnamefont {Ono}},\
  }\href {\doibase 10.7566/JPSJ.85.024701} {\bibfield  {journal} {\bibinfo
  {journal} {Journal of the Physical Society of Japan}\ }\textbf {\bibinfo
  {volume} {85}},\ \bibinfo {pages} {024701} (\bibinfo {year} {2016})},\
  \Eprint {http://arxiv.org/abs/https://doi.org/10.7566/JPSJ.85.024701}
  {https://doi.org/10.7566/JPSJ.85.024701} \BibitemShut {NoStop}%
\bibitem [{\citenamefont {Iwase}\ \emph {et~al.}(2017)\citenamefont {Iwase},
  \citenamefont {Kirkham},\ and\ \citenamefont {Ono}}]{Ono2}%
  \BibitemOpen
  \bibfield  {author} {\bibinfo {author} {\bibfnamefont {S.}~\bibnamefont
  {Iwase}}, \bibinfo {author} {\bibfnamefont {C.~J.}\ \bibnamefont {Kirkham}},
  \ and\ \bibinfo {author} {\bibfnamefont {T.}~\bibnamefont {Ono}},\ }\href
  {\doibase 10.1103/PhysRevB.95.041302} {\bibfield  {journal} {\bibinfo
  {journal} {Phys. Rev. B}\ }\textbf {\bibinfo {volume} {95}},\ \bibinfo
  {pages} {041302} (\bibinfo {year} {2017})}\BibitemShut {NoStop}%
\bibitem [{\citenamefont {Matsushita}\ and\ \citenamefont
  {Oshiyama}(2017)}]{Matsushita_Dit}%
  \BibitemOpen
  \bibfield  {author} {\bibinfo {author} {\bibfnamefont {Y.-i.}\ \bibnamefont
  {Matsushita}}\ and\ \bibinfo {author} {\bibfnamefont {A.}~\bibnamefont
  {Oshiyama}},\ }\href {\doibase 10.1021/acs.nanolett.7b03490} {\bibfield
  {journal} {\bibinfo  {journal} {Nano Letters}\ }\textbf {\bibinfo {volume}
  {17}},\ \bibinfo {pages} {6458} (\bibinfo {year} {2017})},\ \bibinfo {note}
  {pMID: 28898089},\ \Eprint
  {http://arxiv.org/abs/https://doi.org/10.1021/acs.nanolett.7b03490}
  {https://doi.org/10.1021/acs.nanolett.7b03490} \BibitemShut {NoStop}%
\bibitem [{Hij()}]{Hijikata}%
  \BibitemOpen
  \href@noop {} {}\bibinfo {note} {Y. Hijikata, H. Yaguchi, and S. Yoshida,
  Appl. Phys. Exp., {\bf 2}, 021203 (2009).}\BibitemShut {Stop}%
\bibitem [{\citenamefont {Goto}\ and\ \citenamefont
  {Hijikata}(2016)}]{Hijikata1}%
  \BibitemOpen
  \bibfield  {author} {\bibinfo {author} {\bibfnamefont {D.}~\bibnamefont
  {Goto}}\ and\ \bibinfo {author} {\bibfnamefont {Y.}~\bibnamefont
  {Hijikata}},\ }\href {http://stacks.iop.org/0022-3727/49/i=22/a=225103}
  {\bibfield  {journal} {\bibinfo  {journal} {Journal of Physics D: Applied
  Physics}\ }\textbf {\bibinfo {volume} {49}},\ \bibinfo {pages} {225103}
  (\bibinfo {year} {2016})}\BibitemShut {NoStop}%
\bibitem [{\citenamefont {Hirai}\ and\ \citenamefont {Kita}(2017)}]{Hirai}%
  \BibitemOpen
  \bibfield  {author} {\bibinfo {author} {\bibfnamefont {H.}~\bibnamefont
  {Hirai}}\ and\ \bibinfo {author} {\bibfnamefont {K.}~\bibnamefont {Kita}},\
  }\href {\doibase 10.1063/1.4980093} {\bibfield  {journal} {\bibinfo
  {journal} {Applied Physics Letters}\ }\textbf {\bibinfo {volume} {110}},\
  \bibinfo {pages} {152104} (\bibinfo {year} {2017})},\ \Eprint
  {http://arxiv.org/abs/https://doi.org/10.1063/1.4980093}
  {https://doi.org/10.1063/1.4980093} \BibitemShut {NoStop}%
\bibitem [{\citenamefont {Afanasev}\ \emph {et~al.}(1997)\citenamefont
  {Afanasev}, \citenamefont {Bassler}, \citenamefont {Pensl},\ and\
  \citenamefont {Schulz}}]{Afanasev}%
  \BibitemOpen
  \bibfield  {author} {\bibinfo {author} {\bibfnamefont {V.~V.}\ \bibnamefont
  {Afanasev}}, \bibinfo {author} {\bibfnamefont {M.}~\bibnamefont {Bassler}},
  \bibinfo {author} {\bibfnamefont {G.}~\bibnamefont {Pensl}}, \ and\ \bibinfo
  {author} {\bibfnamefont {M.}~\bibnamefont {Schulz}},\ }\href {\doibase
  10.1002/1521-396X(199707)162:1<321::AID-PSSA321>3.0.CO;2-F} {\bibfield
  {journal} {\bibinfo  {journal} {physica status solidi (a)}\ }\textbf
  {\bibinfo {volume} {162}},\ \bibinfo {pages} {321} (\bibinfo {year}
  {1997})}\BibitemShut {NoStop}%
\bibitem [{\citenamefont {Kobayashi}\ \emph {et~al.}(2016)\citenamefont
  {Kobayashi}, \citenamefont {Nakazawa}, \citenamefont {Okuda}, \citenamefont
  {Suda},\ and\ \citenamefont {Kimoto}}]{Kobayashi}%
  \BibitemOpen
  \bibfield  {author} {\bibinfo {author} {\bibfnamefont {T.}~\bibnamefont
  {Kobayashi}}, \bibinfo {author} {\bibfnamefont {S.}~\bibnamefont {Nakazawa}},
  \bibinfo {author} {\bibfnamefont {T.}~\bibnamefont {Okuda}}, \bibinfo
  {author} {\bibfnamefont {J.}~\bibnamefont {Suda}}, \ and\ \bibinfo {author}
  {\bibfnamefont {T.}~\bibnamefont {Kimoto}},\ }\href
  {http://dx.doi.org/10.1063/1.4946863} {\bibfield  {journal} {\bibinfo
  {journal} {Applied Physics Letters}\ }\textbf {\bibinfo {volume} {108}},\
  \bibinfo {pages} {152108} (\bibinfo {year} {2016})}\BibitemShut {NoStop}%
\bibitem [{\citenamefont {Kobayashi}\ \emph {et~al.}(2017)\citenamefont
  {Kobayashi}, \citenamefont {Matsushita}, \citenamefont {Okuda}, \citenamefont
  {Kimoto},\ and\ \citenamefont {Oshiyama}}]{Kobayashi2}%
  \BibitemOpen
  \bibfield  {author} {\bibinfo {author} {\bibfnamefont {T.}~\bibnamefont
  {Kobayashi}}, \bibinfo {author} {\bibfnamefont {Y.-i.}\ \bibnamefont
  {Matsushita}}, \bibinfo {author} {\bibfnamefont {T.}~\bibnamefont {Okuda}},
  \bibinfo {author} {\bibfnamefont {T.}~\bibnamefont {Kimoto}}, \ and\ \bibinfo
  {author} {\bibfnamefont {A.}~\bibnamefont {Oshiyama}},\ }\href@noop {}
  {\bibfield  {journal} {\bibinfo  {journal} {arXiv}\ ,\ \bibinfo {pages}
  {1703.08063}} (\bibinfo {year} {2017})}\BibitemShut {NoStop}%
\bibitem [{\citenamefont {Car}\ and\ \citenamefont {Parrinello}(1985)}]{CPMD}%
  \BibitemOpen
  \bibfield  {author} {\bibinfo {author} {\bibfnamefont {R.}~\bibnamefont
  {Car}}\ and\ \bibinfo {author} {\bibfnamefont {M.}~\bibnamefont
  {Parrinello}},\ }\href@noop {} {\bibfield  {journal} {\bibinfo  {journal}
  {Phys. Rev. Lett.}\ }\textbf {\bibinfo {volume} {55}},\ \bibinfo {pages}
  {2471} (\bibinfo {year} {1985})}\BibitemShut {NoStop}%
\bibitem [{\citenamefont {Iwata}\ \emph {et~al.}(2010)\citenamefont {Iwata},
  \citenamefont {Takahashi}, \citenamefont {Oshiyama}, \citenamefont {Boku},
  \citenamefont {Shiraishi}, \citenamefont {Okada},\ and\ \citenamefont
  {Yabana}}]{RSCPMD}%
  \BibitemOpen
  \bibfield  {author} {\bibinfo {author} {\bibfnamefont {J.~I.}\ \bibnamefont
  {Iwata}}, \bibinfo {author} {\bibfnamefont {D.}~\bibnamefont {Takahashi}},
  \bibinfo {author} {\bibfnamefont {A.}~\bibnamefont {Oshiyama}}, \bibinfo
  {author} {\bibfnamefont {B.}~\bibnamefont {Boku}}, \bibinfo {author}
  {\bibfnamefont {K.}~\bibnamefont {Shiraishi}}, \bibinfo {author}
  {\bibfnamefont {S.}~\bibnamefont {Okada}}, \ and\ \bibinfo {author}
  {\bibfnamefont {K.}~\bibnamefont {Yabana}},\ }\href@noop {} {\bibfield
  {journal} {\bibinfo  {journal} {J. Comp. Phys.}\ }\textbf {\bibinfo {volume}
  {229}},\ \bibinfo {pages} {2339} (\bibinfo {year} {2010})}\BibitemShut
  {NoStop}%
\bibitem [{\citenamefont {Kohn}\ and\ \citenamefont {Sham}(1965)}]{HK}%
  \BibitemOpen
  \bibfield  {author} {\bibinfo {author} {\bibfnamefont {W.}~\bibnamefont
  {Kohn}}\ and\ \bibinfo {author} {\bibfnamefont {L.~J.}\ \bibnamefont
  {Sham}},\ }\href {\doibase 10.1103/PhysRev.140.A1133} {\bibfield  {journal}
  {\bibinfo  {journal} {Phys. Rev.}\ }\textbf {\bibinfo {volume} {140}},\
  \bibinfo {pages} {A1133} (\bibinfo {year} {1965})}\BibitemShut {NoStop}%
\bibitem [{\citenamefont {Hasegawa}\ \emph {et~al.}(2014)\citenamefont
  {Hasegawa}, \citenamefont {Iwata}, \citenamefont {Tsuji}, \citenamefont
  {Takahashi}, \citenamefont {Oshiyama}, \citenamefont {Minami}, \citenamefont
  {Boku}, \citenamefont {Inoue}, \citenamefont {Kitazawa}, \citenamefont
  {Miyoshi},\ and\ \citenamefont {Yokokawa}}]{RSCPMD1}%
  \BibitemOpen
  \bibfield  {author} {\bibinfo {author} {\bibfnamefont {Y.}~\bibnamefont
  {Hasegawa}}, \bibinfo {author} {\bibfnamefont {J.~I.}\ \bibnamefont {Iwata}},
  \bibinfo {author} {\bibfnamefont {M.}~\bibnamefont {Tsuji}}, \bibinfo
  {author} {\bibfnamefont {D.}~\bibnamefont {Takahashi}}, \bibinfo {author}
  {\bibfnamefont {A.}~\bibnamefont {Oshiyama}}, \bibinfo {author}
  {\bibfnamefont {K.}~\bibnamefont {Minami}}, \bibinfo {author} {\bibfnamefont
  {T.}~\bibnamefont {Boku}}, \bibinfo {author} {\bibfnamefont {H.}~\bibnamefont
  {Inoue}}, \bibinfo {author} {\bibfnamefont {Y.}~\bibnamefont {Kitazawa}},
  \bibinfo {author} {\bibfnamefont {I.}~\bibnamefont {Miyoshi}}, \ and\
  \bibinfo {author} {\bibfnamefont {M.}~\bibnamefont {Yokokawa}},\ }\href@noop
  {} {\bibfield  {journal} {\bibinfo  {journal} {International Journal High
  Performance Computing Applications}\ }\textbf {\bibinfo {volume} {28}},\
  \bibinfo {pages} {335} (\bibinfo {year} {2014})}\BibitemShut {NoStop}%
\bibitem [{RSC()}]{RSCPMD2}%
  \BibitemOpen
  \href@noop {} {}\bibinfo {note} {J.-I Iwata,
  https://github.com/j-iwata/RSDFT.}\BibitemShut {Stop}%
\bibitem [{PBE()}]{PBE}%
  \BibitemOpen
  \href@noop {} {}\bibinfo {note} {J. P. Perdew, K. Burke, and M. Ernzerhof,
  Phys. Rev. Lett., {\bf 77}, 3865 (1996).}\BibitemShut {Stop}%
\bibitem [{Nos()}]{Nose}%
  \BibitemOpen
  \href@noop {} {}\bibinfo {note} {S. Nose, J. Chem. Phys., {\bf 81}, 511
  (1984) ; W. G. Hoover, Phys. Rev. A,{\bf 31}, 1695 (1985).}\BibitemShut
  {Stop}%
\bibitem [{\citenamefont {Car}\ \emph {et~al.}(1984)\citenamefont {Car},
  \citenamefont {Kelly}, \citenamefont {Oshiyama},\ and\ \citenamefont
  {Pantelides}}]{car}%
  \BibitemOpen
  \bibfield  {author} {\bibinfo {author} {\bibfnamefont {R.}~\bibnamefont
  {Car}}, \bibinfo {author} {\bibfnamefont {P.~J.}\ \bibnamefont {Kelly}},
  \bibinfo {author} {\bibfnamefont {A.}~\bibnamefont {Oshiyama}}, \ and\
  \bibinfo {author} {\bibfnamefont {S.~T.}\ \bibnamefont {Pantelides}},\ }\href
  {\doibase 10.1103/PhysRevLett.52.1814} {\bibfield  {journal} {\bibinfo
  {journal} {Phys. Rev. Lett.}\ }\textbf {\bibinfo {volume} {52}},\ \bibinfo
  {pages} {1814} (\bibinfo {year} {1984})}\BibitemShut {NoStop}%
\bibitem [{\citenamefont {Kumagai}\ and\ \citenamefont {Oba}(2014)}]{Kumagai}%
  \BibitemOpen
  \bibfield  {author} {\bibinfo {author} {\bibfnamefont {Y.}~\bibnamefont
  {Kumagai}}\ and\ \bibinfo {author} {\bibfnamefont {F.}~\bibnamefont {Oba}},\
  }\href {\doibase 10.1103/PhysRevB.89.195205} {\bibfield  {journal} {\bibinfo
  {journal} {Phys. Rev. B}\ }\textbf {\bibinfo {volume} {89}},\ \bibinfo
  {pages} {195205} (\bibinfo {year} {2014})}\BibitemShut {NoStop}%
\bibitem [{vas()}]{vasp}%
  \BibitemOpen
  \href@noop {} {}\bibinfo {note} {G. Kresse, and J. Furthmuller, Phys. Rev. B,
  {\bf 54}, 11169 (1996). G. Kresse, and D. Joubert, Phys. Rev. B, {\bf 59},
  1758 (1999).}\BibitemShut {Stop}%
\bibitem [{PAW()}]{PAW}%
  \BibitemOpen
  \href@noop {} {}\bibinfo {note} {P. E. Blochl, Phys. Rev. B, {\bf 50}, 17953
  (1994).}\BibitemShut {Stop}%
\bibitem [{Hyd()}]{Hyde}%
  \BibitemOpen
  \href@noop {} {}\bibinfo {note} {J. Hyde, G. E. Scuseria, and M. Ernzerhof,
  J. Chem. Phys. {\bf 118}, 8207 (2003); {\bf 124}, 219906(E)
  (2006).}\BibitemShut {Stop}%
\bibitem [{Kre()}]{Kresse}%
  \BibitemOpen
  \href@noop {} {}\bibinfo {note} {J. Paier, M. Marsman, K. Hummer, G. Kresse,
  I. C. Gerber, and J. G. Angyan, J. Chem. Phys. {\bf 124}, 154709
  (2006).}\BibitemShut {Stop}%
\bibitem [{mat({\natexlab{a}})}]{matsushita}%
  \BibitemOpen
  \href@noop {} {} ({\natexlab{a}}),\ \bibinfo {note} {y. -i. Matsushita, K.
  Nakamura and A. Oshiyama, Phys. Rev. B {\bf 84}, 075205 (2011) and references
  therein.}\BibitemShut {Stop}%
\bibitem [{\citenamefont {Watanabe}\ and\ \citenamefont
  {Hosoi}(2012)}]{offset}%
  \BibitemOpen
  \bibfield  {author} {\bibinfo {author} {\bibfnamefont {H.}~\bibnamefont
  {Watanabe}}\ and\ \bibinfo {author} {\bibfnamefont {T.}~\bibnamefont
  {Hosoi}},\ }in\ \href {\doibase 10.5772/51514} {\emph {\bibinfo {booktitle}
  {Physics and Technology of Silicon Carbide Devices}}},\ \bibinfo {editor}
  {edited by\ \bibinfo {editor} {\bibfnamefont {Y.}~\bibnamefont {Hijikata}}}\
  (\bibinfo  {publisher} {InTech},\ \bibinfo {address} {Rijeka},\ \bibinfo
  {year} {2012})\ Chap.~\bibinfo {chapter} {09}\BibitemShut {NoStop}%
\bibitem [{\citenamefont {Sprik}\ and\ \citenamefont {Ciccotti}(1998)}]{BM}%
  \BibitemOpen
  \bibfield  {author} {\bibinfo {author} {\bibfnamefont {M.}~\bibnamefont
  {Sprik}}\ and\ \bibinfo {author} {\bibfnamefont {G.}~\bibnamefont
  {Ciccotti}},\ }\href {\doibase 10.1063/1.477419} {\bibfield  {journal}
  {\bibinfo  {journal} {The Journal of Chemical Physics}\ }\textbf {\bibinfo
  {volume} {109}},\ \bibinfo {pages} {7737} (\bibinfo {year} {1998})},\ \Eprint
  {http://arxiv.org/abs/https://doi.org/10.1063/1.477419}
  {https://doi.org/10.1063/1.477419} \BibitemShut {NoStop}%
\bibitem [{mat({\natexlab{b}})}]{matsushita_arXiv}%
  \BibitemOpen
  \href@noop {} {} ({\natexlab{b}}),\ \bibinfo {note} {y.-i. Matsushita and A.
  Oshiyama, arXiv:1612.00189 (2016).}\BibitemShut {Stop}%
\bibitem [{\citenamefont {Ito}\ \emph {et~al.}(2015)\citenamefont {Ito},
  \citenamefont {Akiyama}, \citenamefont {Nakamura}, \citenamefont {Ito},
  \citenamefont {Kageshima}, \citenamefont {Uematsu},\ and\ \citenamefont
  {Shiraishi}}]{Akiyama}%
  \BibitemOpen
  \bibfield  {author} {\bibinfo {author} {\bibfnamefont {A.}~\bibnamefont
  {Ito}}, \bibinfo {author} {\bibfnamefont {T.}~\bibnamefont {Akiyama}},
  \bibinfo {author} {\bibfnamefont {K.}~\bibnamefont {Nakamura}}, \bibinfo
  {author} {\bibfnamefont {T.}~\bibnamefont {Ito}}, \bibinfo {author}
  {\bibfnamefont {H.}~\bibnamefont {Kageshima}}, \bibinfo {author}
  {\bibfnamefont {M.}~\bibnamefont {Uematsu}}, \ and\ \bibinfo {author}
  {\bibfnamefont {K.}~\bibnamefont {Shiraishi}},\ }\href
  {http://stacks.iop.org/1347-4065/54/i=10/a=101301} {\bibfield  {journal}
  {\bibinfo  {journal} {Japanese Journal of Applied Physics}\ }\textbf
  {\bibinfo {volume} {54}},\ \bibinfo {pages} {101301} (\bibinfo {year}
  {2015})}\BibitemShut {NoStop}%
\end{thebibliography}%

\onecolumngrid

\end{document}